\documentclass[eps,prd,twocolumn,nofootinbib,showpacs]{revtex4}

\usepackage{graphicx,color}
\usepackage[colorlinks=true,linkcolor=blue,citecolor=blue,urlcolor=blue]{hyperref}

\begin{document}

\title{Revisiting the top-quark pair production at future $e^+e^-$ colliders }

\author{Jin Ma$^1$}

\author{Sheng-Quan Wang$^1$}
\email[email:]{sqwang@cqu.edu.cn}

\author{Ting Sun$^1$}

\author{Jian-Ming Shen$^{2}$}

\author{Xing-Gang Wu$^3$}
\email[email:]{wuxg@cqu.edu.cn}

\address{$^1$Department of Physics, Guizhou Minzu University, Guiyang 550025, P.R. China}
\address{$^2$School of Physics and Electronics, Hunan Provincial Key Laboratory of High-Energy Scale Physics and Applications, Hunan University, Changsha 410082, P.R. China}
\address{$^3$Department of Physics, Chongqing Key Laboratory for Strongly Coupled Physics, Chongqing University, Chongqing 401331, P.R. China}

\date{\today}

\begin{abstract}

In this paper, we reanalyze the top-quark pair production at the next-to-next-to-leading order (NNLO) in QCD at future $e^+e^-$ colliders by using the Principle of Maximum Conformality (PMC) method. The PMC renormalization scales in $\alpha_s$ are determined by absorbing the non-conformal $\beta$ terms by recursively using the Renormalization Group Equation (RGE). Unlike the conventional scale-setting method of fixing the scale at the center-of-mass energy $\mu_r=\sqrt{s}$, the determined PMC scale $Q_\star$ is far smaller than the $\sqrt{s}$ and increases with the $\sqrt{s}$, yielding the correct physical behavior for the top-quark pair production process. Moreover, the convergence of the pQCD series for the top-quark pair production is greatly improved due to the elimination of the renormalon divergence. For a typical collision energy of $\sqrt{s}=500$ GeV, the PMC scale is $Q_\star=107$ GeV; the QCD correction factor $K$ for conventional results is $K\sim1+0.1244^{+0.0102+0.0012}_{-0.0087-0.0011}+0.0184^{-0.0086+0.0002}_{+0.0061-0.0003}$, where the first error is caused by varying the scale $\mu_r\in[\sqrt{s}/2, 2\sqrt{s}]$ and the second error is from the top-quark mass $\Delta{m_t}=\pm0.7$ GeV. After applying the PMC, the renormalization scale uncertainty is eliminated and the QCD correction factor $K$ is improved to $K\sim 1+0.1507^{+0.0015}_{-0.0015}-0.0057^{+0.0001}_{-0.0000}$, where the error is from the top-quark mass $\Delta{m_t}=\pm0.7$ GeV. The PMC improved predictions for the top-quark pair production are helpful for detailed studies of properties of the top-quark at future $e^+e^-$ colliders.

\end{abstract}

\maketitle

\section{Introduction}
\label{sec:1}

The top-quark is the heaviest elementary particle in the Standard Model (SM). The detailed study of top-quark properties is of great significance for testing the SM and new physics beyond the SM. Experimentally, the top-quark has been extensively studied at hadron colliders since it was discovered by the CDF and D0 Collaborations~\cite{CDF:1995wbb, D0:1995jca}. However, the hadron collider platform has an unclean experimental environment, which reduces the precision of experimental measurements. Currently, the uncertainties are about $0.3\%$~\cite{ATLAS:2018fwq, CMS:2015lbj} for the top-quark mass and $3$-$4\%$~\cite{ATLAS:2019hau, CMS:2018fks} for the top-quark pair production cross-section measured by the ATLAS and CMS Collaborations at the Large Hadron Collider(LHC).

The study of top-quark properties is among the core physics issues at future lepton colliders~\cite{ILC:2013jhg, Bambade:2019fyw, Vos:2016til, CEPCStudyGroup:2018ghi, CLICdp:2018esa, FCC:2018evy, ILDConceptGroup:2020sfq}. Compared to hadron colliders, lepton colliders have a clean collision environment, which will reach a new level of precision for measurements of top-quark properties. For example, at the International Linear Collider~\cite{ILDConceptGroup:2020sfq}, the uncertainty can be improved to $50$ MeV for the top-quark mass measured from the top-quark pair production near threshold. The precision of experimental measurements at future lepton colliders calls for precise theoretical predictions for top-quark properties.

Significant efforts have been made to investigate the production of the top-quark pair at $e^+e^-$ colliders. QCD corrections for the top-quark pair production at the threshold have been calculated at next-to-next-to-leading order (NNLO)~\cite{Czarnecki:1997vz, Beneke:1997jm, Hoang:1998xf, Beneke:1999qg}, as well as next-to-next-to-next-to-leading order (NNNLO) QCD calculations have been given in Refs.\cite{Beneke:2015kwa, Beneke:2016kkb}. For the top-quark pair production in the continuum, QCD corrections at next-to-leading order (NLO) have been known for a long time~\cite{Jersak:1981sp}. Also NLO electroweak corrections are given in Refs.\cite{Beenakker:1991ca, Fleischer:2003kk,Hahn:2003ab, Khiem:2012bp}, and NLO QCD predictions for off-shell top-quark pair production and decay was presented in Ref.\cite{ChokoufeNejad:2016qux}. The total cross-sections were computed at higher orders in high energy expansion or by using Pad$\acute{e}$ approximation~\cite{SGG1986, Chetyrkin:1990kr, Chetyrkin:1994ex, Harlander:1997kw, Chetyrkin:1997qi, Chetyrkin:1996cf, Chetyrkin:1997pn, Kiyo:2009gb}. The computation of the cross-section and differential distributions at NNLO QCD with full top-mass dependence were reported in Refs.~\cite{Gao:2014nva, Gao:2014eea, Chen:2016zbz}, and NNLO QCD corrections to the top-quark pair productions with polarized electron and positron beams are recently derived in Ref.\cite{Bernreuther:2023jgp}. NNNLO QCD corrections to the total cross-section was recently investigated in Ref.\cite{Chen:2022vzo}.

It is well known that fixed-order perturbative QCD (pQCD) predictions suffer from the ambiguity of the renormalization scale. The conventional scale-setting method usually assumes the renormalization scale as the typical momentum flow of the process, and theoretical uncertainties are estimated by varying the scale over an arbitrary range. For the calculation of the top-quark pair production at $e^+e^-$ colliders, the renormalization scale is usually chosen as the center-of-mass energy $\sqrt{s}$, then varying the scale within $[\sqrt{s}/2, 2\sqrt{s}]$ to ascertain its uncertainty. By using the conventional scale-setting method, the top-quark pair production cross-sections are plagued by the large renormalization scale uncertainty. Moreover, estimating unknown higher-order QCD corrections is unreliable since the NNLO result do not overlap with the NLO prediction obtained by varying the scale $\mu_r\in[\sqrt{s}/2,2\sqrt{s}]$. As a matter of fact that we actually do not know what is the correct range of variation of the scale in order to have reliable quantitative predictions for theoretical uncertainties.

It is important to find a correct method to solve the renormalization scale ambiguity and then to achieve a precise prediction for the top-quark pair production at $e^+e^-$ colliders. The Principle of Maximum Conformality (PMC)~\cite{Brodsky:2011ta, Brodsky:2012rj, Brodsky:2011ig, Mojaza:2012mf, Brodsky:2013vpa} provides a systematic way to eliminate renormalization scheme and scale ambiguities in pQCD predictions. The PMC method extends the Brodsky-Lepage-Mackenzie (BLM) scale-setting method~\cite{Brodsky:1982gc} to all orders, and it reduces in the Abelian limit to the Gell-Mann-Low method~\cite{GellMann:1954fq}. The QCD predictions using PMC do not depend on the renormalization scheme, satisfying the principles of Renormalization Group invariance (RGI)~\cite{Brodsky:2012ms, Wu:2014iba, Wu:2018cmb, Wu:2019mky}. The PMC method has been successfully applied to many pQCD physical processes (see e.g.,~\cite{Wang:2023ttk, Wu:2015rga} for reviews). Due to the renormalon divergence disappears in the pQCD series, a more convergent perturbative series can be in general achieved.

By applying the PMC method to the top-quark pair hadroproduction processes, the resulting production cross-sections agree with precise experimental data, and the large discrepancies of the top-quark forward-backward asymmetries between QCD estimations and experimental measurements are greatly reduced~\cite{Brodsky:2012rj, Brodsky:2012sz, Brodsky:2012ik, Wang:2014sua, Wang:2015lna}. It is noted that after achieving higher-order pQCD corrections, the conventional predictions are also in agreement with experimental measurements~\cite{Czakon:2013goa, Czakon:2015owf, Czakon:2017lgo}. A precise top-quark mass can be extracted by the comparison of PMC predictions of production cross-sections with experimental data at the LHC~\cite{Wang:2017kyd, Wang:2020mel}. Recently, a improved QCD prediction for the top-quark decay process obtained by using the PMC is also given in Ref.\cite{Meng:2022htg}. In addition, the PMC scale-setting method provides a self-consistent analysis and yields the correct physical behavior of the scale for the top-quark pair production at the threshold at $e^+e^-$ colliders~\cite{Brodsky:1995ds, Wang:2020ckr}.

Achieving a reliable and precise prediction for the top-quark pair production at future $e^+e^-$ colliders is desirable. In this paper, as a step forward of our previous calculations for the top-quark production at the threshold at NNLO~\cite{Wang:2020ckr}, we will calculate the top-quark pair production beyond the threshold region at future $e^+e^-$ colliders by applying the PMC method~\footnote{In this paper, we calculate the top-quark pair production at future $e^+e^-$ colliders, where the collision energy $\sqrt{s}$ is far from the threshold region. A detailed PMC analysis for the top-quark pair production at the threshold region at NNNLO is given in Ref.\cite{Yan:2023mjj}.}.

The remaining sections of this paper are organized as follows. We present in section~\ref{sec:2} our calculation technology for applying the PMC method to the top-quark pair production at $e^+e^-$ colliders. In section~\ref{sec:3}, we give numerical results and discussions. This paper is concluded with a summary in section~\ref{sec:4}.

\section{PMC scale setting for the $t\bar{t}$ production at future $e^+e^-$ colliders}
\label{sec:2}

QCD corrections for the top-quark pair production cross-section $\sigma^{t\bar{t}}$ at future $e^+e^-$ colliders at the NNLO can be written as
\begin{eqnarray}
\sigma^{t\bar{t}}_{\rm NNLO}&=&\sigma^{t\bar{t}}_{\rm LO}\left[1+\triangle_1+\triangle_2\right],
\label{sigma_NNLO}
\end{eqnarray}
where $\sigma^{t\bar{t}}_{\rm LO}$ is the production cross-section at LO, and the NLO and NNLO QCD correction terms are
\begin{eqnarray}
\triangle_1&=&c_1\,a_s(\mu_r), \\
\triangle_2&=&c_2(\mu_r)\,a^2_s(\mu_r),
\end{eqnarray}
respectively. The coefficient $c_1$ ($c_2$) is the NLO (NNLO) correction coefficient; $a_s(\mu_r)=\alpha_s(\mu_r)/(4\pi)$ is the QCD running coupling constant, $\mu_r$ is the renormalization scale.

The coefficient $c_2$ can be further divided into $n_f$-dependent and $n_f$-independent parts, i.e.,
\begin{eqnarray}
c_2(\mu_r)=c_{2,0}(\mu_r)+c_{2,1}(\mu_r)\,n_{f},
\end{eqnarray}
where $n_f$ is the number of active light quark flavors. By using the relation of $\beta_{0}=11-2/3\,n_f$, the top-quark pair production cross-section $\sigma^{t\bar{t}}$ in Eq.(\ref{sigma_NNLO}) can be rewritten as
\begin{eqnarray}
\sigma^{t\bar{t}}_{\rm NNLO}&=&\sigma^{t\bar{t}}_{\rm LO}\left[1+r_{1,0}\,a_s(\mu_r)+\left(r_{2,0}(\mu_r) \right.\right. \nonumber\\
&&\left.\left.+r_{2,1}(\mu_r)\,\beta_{0}\right)\,a^2_s(\mu_r)\right],
\label{sigma_NNLObe}
\end{eqnarray}
where the coefficients $r_{1,0}$ and $r_{2,0}(\mu_r)$ are conformal coefficients, and the coefficient $r_{2,1}(\mu_r)$ stands for non-conformal coefficient. They are
\begin{eqnarray}
r_{1,0}&=&c_1, \\
r_{2,0}(\mu_r)&=&c_{2,0}(\mu_r)+\frac{33}{2}\,c_{2,1}(\mu_r), \\
r_{2,1}(\mu_r)&=&-\frac{3}{2}\,c_{2,1}(\mu_r).
\end{eqnarray}

By applying the PMC method to heavy quark pair production processes at $e^+e^-$ colliders, two distinctly different scales are in general determined~\cite{Brodsky:1995ds}. For our current analysis of the top-quark pair production beyond the threshold region at future $e^+e^-$ colliders, the center-of-mass energy such as the typical collision energy of $\sqrt{s}=500$ GeV is much larger than the top-quark mass. Consequently, the top-quark velocity is $v\sim1$, and the Coulomb correction is negligible. We thus only need to determine the PMC scale for the hard virtual corrections. It is noted that the good agreements of the top-quark pair production cross-sections in the continuum~\cite{Gao:2014nva, Gao:2014eea, Chen:2016zbz} with ones from high-energy expansions~\cite{Harlander:1997kw, Chetyrkin:1997qi} was demonstrated in Ref.\cite{Gao:2014eea}. We adopt the QCD correction coefficients $c_1$ and $c_2$ from Refs.~\cite{Harlander:1997kw, Chetyrkin:1997qi} for our analysis.

We adopt the PMC single-scale method~\cite{Shen:2017pdu} for our current NNLO QCD analysis, and an overall scale can be determined by eliminating the non-conformal coefficient. The resulting PMC prediction for the top-quark pair production cross-section $\sigma^{t\bar{t}}$ is
\begin{eqnarray}
\sigma^{t\bar{t}}_{\rm NNLO}=\sigma^{t\bar{t}}_{\rm LO}\left[1+r_{1,0}\,a_s(Q_\star)+r_{2,0}(\mu_r)\,a^2_s(Q_\star)\right],
\label{sigma_NNLOpmc}
\end{eqnarray}
where $Q_\star$ is the PMC scale,
\begin{eqnarray}
Q_\star=\mu_r\,\exp\left[-\frac{r_{2,1}(\mu_r)}{2\,r_{\rm 1,0}}\right].
\label{PMCscaleQ}
\end{eqnarray}
After applying the PMC, we can see from Eq.(\ref{sigma_NNLOpmc}) that only the conformal coefficients remains in the pQCD series, and the non-conformal coefficient disappear. It is noted that the PMC scale $Q_\star$ and the conformal coefficient $r_{2,0}$ are independent of the choice of the renormalization scale $\mu_r$, the top-quark pair production cross-section in Eq.(\ref{sigma_NNLOpmc}) eliminates the renormalization scale ambiguity.

\section{Numerical results and discussions}
\label{sec:3}

For the numerical calculations, input parameters are taken from the Particle Data Group (PDG)~\cite{ParticleDataGroup:2022pth}, such as the top-quark pole mass $m_t=172.5$ GeV, the $Z$ boson mass $m_Z=91.1876$ GeV, the $W$ boson mass $m_W=80.377$ GeV, the two-loop $\overline{\rm MS}$ scheme QCD coupling is evaluated from $\alpha_s(m_Z)=0.1179$. The sine of the weak mixing angle is evaluated by $s^2_\theta=(1-m^2_W/m^2_Z)$. The electroweak coupling is fixed by the $G_\mu$ scheme, i.e., $\alpha=\sqrt{2}G_\mu m^2_Ws^2_\theta/\pi$ with $G_\mu=1.166379\times10^{-5}$ GeV$^{-2}$.

\begin{table} [htb]
\begin{tabular}{|c||c|c|c|c|c|}
\hline
~~ ~~  & ~$\mu_r$~ & ~$\sigma^{t\bar{t}}_{\rm LO}$~ & ~NLO~ & ~NNLO~ & ~$\sigma^{t\bar{t}}_{\rm NNLO}$~ \\
\hline
~~ ~~  & ~$\sqrt{s}/2$~ &~0.5512~& ~0.0742~ & ~0.0054~ & ~0.6308~  \\
~Conv.~& ~$\sqrt{s}$~   &~0.5512~& ~0.0686~ & ~0.0101~ & ~0.6299~  \\
~~ ~~  & ~$2\sqrt{s}$~  &~0.5512~& ~0.0638~ & ~0.0135~ & ~0.6285~  \\
\hline
~PMC~ & ~~         & ~0.5512~ & ~0.0831~ & ~-0.0032~  & ~0.6311~  \\
\hline
\end{tabular}
\caption{The top-quark pair production cross-sections (in unit pb) at $e^+e^-$ colliders at $\sqrt{s}=500$ GeV using the conventional (Conv.) and PMC scale settings.
\label{tab1}  }
\end{table}

According to the conventional scale-setting method, the renormalization scale is simply set to the center-of-mass energy $\sqrt{s}$, and then the uncertainty is estimated by varying the scale over an arbitrary range, e.g., $\mu_r\in[\sqrt{s}/2, 2\sqrt{s}]$. In Table \ref{tab1} we present the top-quark pair production cross-sections at $e^+e^-$ colliders at $\sqrt{s}=500$ GeV using the conventional and PMC scale settings. The LO production cross-section does not involve the strong interaction and provides dominant contributions. At present NNLO level, the scale uncertainties are rather large for the QCD correction terms, i.e., the NLO QCD correction term is $0.0686$ pb for $\mu_r=\sqrt{s}$, its scale error is $\sim[-7.0\%,+8.2\%]$ for $\mu_r\in[\sqrt{s}/2, 2\sqrt{s}]$; the NNLO QCD correction term is $0.0101$ pb for $\mu_r=\sqrt{s}$, its scale error is $\sim[+33.2\%, -46.7\%]$ for $\mu_r\in[\sqrt{s}/2, 2\sqrt{s}]$. We thus can not decide what is the exact QCD correction terms for each perturbative order. Table \ref{tab1} shows that the NLO QCD correction term decreases and the NNLO QCD correction term increases with the increasing of the scale $\mu_r$. Thus, the scale uncertainty cancel each other out, leading to a small scale uncertainty $\sim[-1.8\%,+1.1\%]$ for the total QCD correction terms.

\begin{figure} [htb]
\centering
\includegraphics[width=0.40\textwidth]{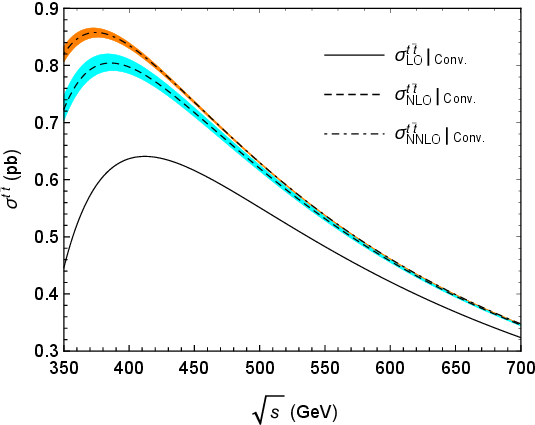}
\caption{The top-quark pair production cross-sections at $e^+e^-$ colliders versus the collision energy $\sqrt{s}$ using the conventional (Conv.) scale setting. The solid, dashed and dotdashed lines represent the conventional predictions at LO, NLO and NNLO, respectively. The bands are caused by varying the scale $\mu_r\in[\sqrt{s}/2,2\sqrt{s}]$. }
\label{ConvVsqS}
\end{figure}

In addition, in the case of conventional scale setting, the total top-quark pair production cross-section at NNLO for $\sqrt{s}=500$ GeV is $\sigma^{t\bar{t}}_{\rm NNLO}=0.6299^{+0.0009}_{-0.0014}$ pb, which does not overlap with the NLO prediction $\sigma^{t\bar{t}}_{\rm NLO}=0.6198^{+0.0048}_{-0.0056}$ pb. More explicitly, in Fig.(\ref{ConvVsqS}) we present the top-quark pair production cross-sections at $e^+e^-$ colliders versus the collision energy $\sqrt{s}$ using the conventional scale setting. It shows that the production cross-section $\sigma^{t\bar{t}}_{\rm NNLO}$ at NNLO does not overlap with the NLO prediction $\sigma^{t\bar{t}}_{\rm NLO}$, specially near the threshold region. Thus estimating unknown higher-order QCD terms by varying the scale $\mu_r\in[\sqrt{s}/2,2\sqrt{s}]$ is unreliable for the top-quark pair production at $e^+e^-$ colliders. In fact, simply varying the renormalization scale is only sensitive to the non-conformal $\beta$ terms, but not to the conformal terms.

In contrast, after using PMC scale setting, the renormalization scale uncertainty is eliminated. The NLO QCD correction term is almost fixed to be $0.0831$ pb, and the NNLO QCD correction term is $-0.0032$ pb for any choice of the scale $\mu_r$. It is noted that the scale-independent conformal coefficient $r_{\rm 2,0}(\mu_r)$ is quite different from the conventional coefficient $c_2(\mu_r)$, i.e., the coefficient $r_{\rm 2,0}(\mu_r)$ is fixed to be
\begin{eqnarray}
r_{\rm 2,0}(\mu_r)=-68.34
\end{eqnarray}
for any choice of the scale $\mu_r$. The conventional coefficient $c_2(\mu_r)$ is
\begin{eqnarray}
c_2(\mu_r)=320.57\pm174.80
\end{eqnarray}
for $\mu_r\in[\sqrt{s}/2,2\sqrt{s}]$. The scale-independent conformal coefficient $r_{\rm 2,0}(\mu_r)$ is negative value and is much smaller than the conventional coefficient $c_2(\mu_r)$. Thus, the NNLO QCD correction term provides a positive contribution using conventional scale setting; it becomes a negative contribution after using the PMC scale setting. The scale-independent total top-quark pair production cross-section at NNLO by using the PMC is $\sigma^{t\bar{t}}_{\rm NNLO}=0.6311$ pb.

A NNNLO QCD correction to this process has been performed in Ref.\cite{Chen:2022vzo}. It has a numerically small QCD correction term, even using conventional scale setting, suggesting improved pQCD convergence at NNNLO. As mentioned above, the scale uncertainties are rather large for the QCD correction terms. For $\sqrt{s}=500$ GeV, the NNNLO QCD correction terms are $-0.0012$, $0.0003$, $0.0019$ pb for the scale $\mu_r=\sqrt{s}/2$, $\sqrt{s}$, $2\sqrt{s}$, respectively. The NNNLO QCD correction term increases with the increasing of the scale $\mu_r$ and is $0.0031$ pb for the scale of $\mu_r=3\sqrt{s}$. Its magnitude is very close to the that of the scale-independent NNLO QCD correction term $-0.0032$ pb obtained by using the PMC.

In the case of conventional scale setting, the QCD correction factor $K=1+\triangle_1+\triangle_2$ is
\begin{eqnarray}
K\sim 1+0.1244+0.0184
\end{eqnarray}
for $\mu_r=\sqrt{s}$. The QCD correction factor $K$ is changed to
\begin{eqnarray}
K\sim 1+0.1346+0.0098
\end{eqnarray}
for $\mu_r=\sqrt{s}/2$. The NLO QCD correction term is increased while the NNLO QCD correction term is suppressed compared to the result obtained at $\mu_r=\sqrt{s}$, which implies that the convergence of the pQCD series is improved. However, the QCD correction factor $K$ is
\begin{eqnarray}
K\sim 1+0.1157+0.0245
\end{eqnarray}
for $\mu_r=2\sqrt{s}$. The NLO QCD correction term is suppressed while the NNLO QCD correction term is increased compared to the case of $\mu_r=\sqrt{s}$, which implies the slow convergence of the pQCD series at $\mu_r=2\sqrt{s}$. Thus, one cannot decide the intrinsic convergence of the pQCD series. The quality of the convergence of the pQCD series depends on the choice of the renormalization scale.

After applying PMC scale setting, the QCD correction factor $K$ is
\begin{eqnarray}
K\sim 1+0.1507-0.0057
\end{eqnarray}
for any choice of the renormalization scale $\mu_r$. Due to the non-conformal $\beta$ terms disappear in the pQCD series, the NLO QCD correction term is greatly increased while the NNLO QCD correction term is highly suppressed compared to the conventional results obtained in the range of $\mu_r\in[\sqrt{s}/2,2\sqrt{s}]$. Thus, the convergence of the pQCD series using PMC scale setting is greatly improved.

\begin{figure} [htb]
\centering
\includegraphics[width=0.40\textwidth]{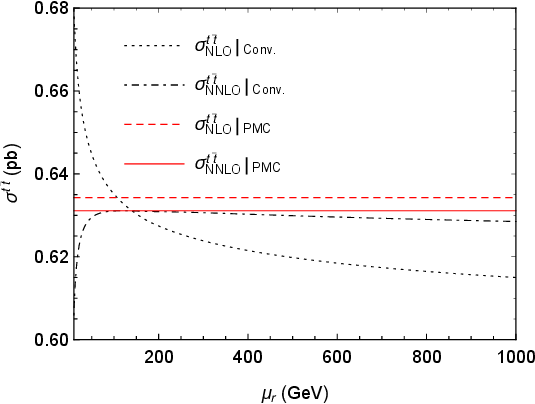}
\caption{The top-quark pair production cross-sections at $e^+e^-$ colliders at $\sqrt{s}=500$ GeV versus the renormalization scale $\mu_r$ at NLO and NNLO using the conventional (Conv.) and PMC scale settings. The dotted and dotdashed lines stand for the conventional predictions at NLO and NNLO, and the dashed and solid lines stand for the PMC predictions at NLO and NNLO, respectively. }
\label{topConPMC}
\end{figure}

In Fig.(\ref{topConPMC}) we present the total top-quark pair production cross-sections at $e^+e^-$ colliders at $\sqrt{s}=500$ GeV versus the renormalization scale $\mu_r$ at NLO and NNLO using the conventional and PMC scale settings. In the case of conventional scale setting, the total top-quark pair production cross-section $\sigma^{t\bar{t}}_{\rm NLO}$ at NLO depends heavily on the choice of the scale $\mu_r$. As mentioned above, the scale dependence cancel out between the NLO and NNLO QCD correction terms. Thus, the scale dependence is decreased by the inclusion of the NNLO QCD correction.

In the case of PMC scale setting, the scale dependence of the QCD correction terms at each perturbative order is eliminated. The resulting total top-quark pair production cross-sections at NLO and NNLO are scale-independent, as show by Fig.(\ref{topConPMC}).

\begin{figure} [htb]
\centering
\includegraphics[width=0.40\textwidth]{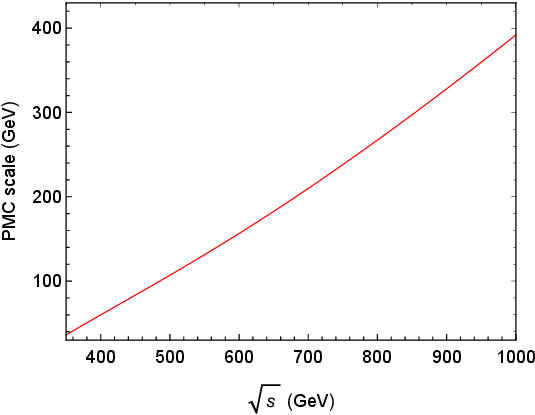}
\caption{The PMC scale $Q_\star$ versus the center-of-mass energy $\sqrt{s}$ for the top-quark pair production at $e^+e^-$ colliders.}
\label{topPMCscale}
\end{figure}

By using the Eq.(\ref{PMCscaleQ}), the PMC scale $Q_\star$ is independent of the choice of the renormalization scale $\mu_r$. It is noted that the determined PMC scale is quite different from the conventional choice of $\mu_r=\sqrt{s}$. For the typical collision energy of $\sqrt{s}=500$ GeV, the PMC scale is
\begin{eqnarray}
Q_\star=0.214\,\sqrt{s}=107\,\, {\rm GeV},
\end{eqnarray}
which is much smaller than the conventional choice $\mu_r=\sqrt{s}=500$ GeV. The PMC scale $Q_\star$ is collision energy $\sqrt{s}$ dependent. More explicitly, we present the PMC scale $Q_\star$ versus the collision energy $\sqrt{s}$ for the top-quark pair production at $e^+e^-$ colliders in Fig.(\ref{topPMCscale}). Figure (\ref{topPMCscale}) shows that the PMC scale $Q_\star$ increases with the collision energy $\sqrt{s}$. The value of the scale $Q_\star$ near the threshold region is much smaller than the top-quark mass $m_t$, yielding large production cross-sections in comparison with the results predicted using conventional scale setting. The dynamics of the PMC scale $Q_\star$ thus signals the correct physical behavior for the top-quark pair production at $e^+e^-$ colliders.

As shown by figure (\ref{topConPMC}), the total top-quark pair production cross-section $\sigma^{t\bar{t}}_{\rm NNLO}$ at NNLO first increases and then decreases with increasing scale $\mu_r$ using conventional scale setting. We obtain the maximum value of the cross-section at a small scale of $\mu_r=120$ GeV. It is noted that near the small scale range of $\mu_r\sim 120$ GeV, the convergence of the pQCD series will be greatly improved, as well as the total top-quark pair production cross-section close to the scale-independent PMC prediction. This indicates that the effective momentum flow for the top-quark pair production at $e^+e^-$ colliders should be $\mu_r\ll \sqrt{s}$, far lower than the conventionally suggested $\mu_r=\sqrt{s}$.

The PMC scales are determined unambiguously by the non-conformal $\beta$ terms, varying the PMC scale would break the renormalization group invariance and then lead to unreliable predictions~\cite{Wu:2014iba}. Thus, varying the scale to estimate the unknown higher-order contributions is not applicable for PMC predictions. The Bayesian-based approach provides a reliable way to estimate the unknown higher-order contributions, which predicts the magnitude of the unknown higher-order contributions based on an optimized analysis of probability density distribution. We adopt the bayesian analysis~\cite{Cacciari:2011ze, Shen:2022nyr, Yan:2022foz} for the estimate of the unknown higher-order contributions. The resulting NNLO PMC results are well within the error bars predicted from the NLO PMC calculations for the top-quark pair production cross-section.

The PMC predictions eliminate the renormalization scale uncertainty, there are uncertainties from other input parameters, such as the top-quark pole mass, the $Z$ boson mass and the $W$ boson mass. In the case of conventional scale setting, the uncertainty caused by the top-quark mass $m_t=172.5\pm0.7$ GeV for the QCD correction factor is $K\sim 1+0.1244^{+0.0012}_{-0.0011}+0.0184^{+0.0002}_{-0.0003}$, which is very small compared to the scale uncertainty caused by $\mu_r\in[\sqrt{s}/2, 2\sqrt{s}]$. After applying the PMC, the uncertainty caused by $m_t=172.5\pm0.7$ GeV for the QCD correction factor is $K\sim 1+0.1507^{+0.0015}_{-0.0015}-0.0057^{+0.0001}_{-0.0000}$. The uncertainties from the $W$ boson mass $m_W=80.377\pm0.012$ GeV and the $Z$ boson mass $m_Z=91.1876\pm0.0021$ GeV are negligible for the QCD correction factor $K$.

Finally, the total top-quark pair production cross-sections at representative $\sqrt{s}$ values are
\begin{eqnarray}
\sigma^{t\bar{t}}_{\rm NNLO}|_{\rm PMC}&=&0.8470\pm0.0089^{+0.0041}_{-0.0042}\pm0.0017 \,\, {\rm pb}, \nonumber
\end{eqnarray}
for $\sqrt{s}=400$ GeV,
\begin{eqnarray}
\sigma^{t\bar{t}}_{\rm NNLO}|_{\rm PMC}&=&0.6311\pm0.0030\pm0.0011\pm0.0006 \,\, {\rm pb}, \nonumber
\end{eqnarray}
for $\sqrt{s}=500$ GeV,
\begin{eqnarray}
\sigma^{t\bar{t}}_{\rm NNLO}|_{\rm PMC}&=&0.4615\pm0.0015\pm0.0004\pm0.0003 \,\, {\rm pb}, \nonumber
\end{eqnarray}
for $\sqrt{s}=600$ GeV,
\begin{eqnarray}
\sigma^{t\bar{t}}_{\rm NNLO}|_{\rm PMC}&=&0.3474\pm0.0010\pm0.0002\pm0.0002 \,\, {\rm pb}, \nonumber
\end{eqnarray}
for $\sqrt{s}=700$ GeV. The first errors are from the unknown higher-order contributions estimated by using the bayesian analysis \cite{Cacciari:2011ze, Shen:2022nyr, Yan:2022foz}, the second errors are caused by the top-quark mass $\Delta{m_t}=\pm0.7$ GeV, and the third errors are from the coupling constant $\Delta \alpha_s(m_Z)=\pm0.0009$. Due to the top-quark pair production cross-section eliminates the renormalization scale ambiguity, as well as the convergence of the pQCD series is greatly improved, the PMC improved predictions for the top-quark pair productions are helpful for detailed studies of properties of the top-quark at future $e^+e^-$ colliders.

\section{Summary }
\label{sec:4}

In the case of conventional scale-setting, the renormalization scale uncertainty becomes one of the most important errors in pQCD predictions. For the top-quark pair production at $e^+e^-$ colliders, the scale uncertainties are rather large for the QCD correction; the quality of the convergence of the pQCD series depends on the choice of the renormalization scale. Thus, one can not decide what is the exact QCD correction terms for each perturbative order and can not decide the intrinsic convergence of the pQCD series. In addition, estimating unknown higher-order QCD terms by varying the scale $\mu_r\in[\sqrt{s}/2,2\sqrt{s}]$ is unreliable for the top-quark pair production at $e^+e^-$ colliders, since the production cross-section at NNLO does not overlap with the NLO prediction.

After using PMC scale setting, the scale in $\alpha_s$ are determined by absorbing the non-conformal $\beta$ terms, and the resulting PMC scale $Q_\star$ is far smaller than the $\sqrt{s}$ and increases with the $\sqrt{s}$, reflecting the increasing virtuality of the QCD dynamics and yielding the correct physical behavior for the top-quark pair production process. The predicted top-quark pair production cross-section eliminates the renormalization scale ambiguity. Moreover, the convergence of the pQCD series is greatly improved due to the elimination of the renormalon divergence. For a typical collision energy of $\sqrt{s}=500$ GeV, the QCD correction factor is improved to $K\sim 1+0.1507-0.0057$, and the predicted production cross-section is
\begin{eqnarray}
\sigma^{t\bar{t}}_{\rm NNLO}|_{\rm PMC}=0.6311\pm0.0033\,\, {\rm pb},
\end{eqnarray}
Our PMC improved predictions for the top-quark pair production are helpful for detailed studies of properties of the top-quark at future $e^+e^-$ colliders.

\hspace{1cm}

{\bf Acknowledgements}: This work was supported in part by the Natural Science Foundation of China under Grants No.12175025, No.12147102 and No.12265011; by the Projects of Guizhou Provincial Department under Grants No.YQK[2023]016, No.ZK[2023]141, No.[2020]1Y027 and No.GZMUZK[2022]PT01.

\end{document}